\documentclass{llncs}
\usepackage{stmaryrd}
\usepackage{hyperref}
\usepackage{mathtools}
\usepackage{enumerate}  
\usepackage{url}
\usepackage[pdftex]{graphicx}
\usepackage{amsmath}
\usepackage{amsfonts}
\usepackage{galois}
\usepackage{mathabx}
\usepackage{todonotes}
\usepackage{paralist}
\usepackage{multirow}
\usepackage[font=small,skip=0pt]{caption}
\usepackage{subfigure}
\usepackage{algorithmicx}
\usepackage[noend]{algpseudocode}
\usepackage{algorithm}
\usepackage{fancyvrb}
\usepackage{lscape}
\usepackage{listings}
\usepackage{pdflscape}
\usepackage{wrapfig}
\usepackage{lipsum}
\usepackage{etoolbox}
\usepackage{tikz}
\usetikzlibrary{shapes}
\usetikzlibrary{arrows}
\usetikzlibrary{positioning}
\usetikzlibrary{automata}
\usetikzlibrary{calc}

\usepackage{pifont}
\newcommand{\cmark}{\ding{51}}%
\newcommand{\xmark}{\ding{55}}%

\IfFileExists{./tmp/rm-margin.tex}{\input{tmp/rm-margin}}{}

\makeatletter
\newcommand{\dashedrightarrow}[1][2pt]{%
  \settowidth{\@tempdima}{$\rightarrow$}\rightarrow
  \makebox[-\@tempdima]{\hskip-1.5ex\color{white}\rule[0.5ex]{#1}{1pt}}
  \phantom{\rightarrow}
}
\makeatother

\newcommand{\integers}{\mathbb{Z}}

\newcommand{\defeq}{\triangleq}

\newcommand{\ppre}{\ensuremath{\mathsf{Pre}}}
\newcommand{\ppost}{\ensuremath{\mathsf{Post}}}
\newcommand{\mE}{\ensuremath{\mathcal{E}}}
\newcommand{\mexit}[2]{\ensuremath{({#1} < {\mE}_{#2})}}
\newcommand{\ltrue}{\mathbf{tt}}
\newcommand{\lfalse}{\mathbf{ff}}
\newcommand{\limplies}{\Rightarrow}

\newcommand{\ourtool}{\textsc{Tiler}}
\newcommand{\zthree}{\textsc{Z3}}

\newcommand{\cbmc}{\textsc{CBMC}}
\newcommand{\booster}{\textsc{Booster}}
\newcommand{\smackpluscorral}{\textsc{SMACK+Corral}}
\newcommand{\corral}{\textsc{Corral}}

\newcommand{\invgen}{\textsc{InvGen}}

\newcommand{\daikon}{\textsc{Daikon}}
\newcommand{\vaphor}{\textsc{Vaphor}}
\newcommand{\spacer}{\textsc{Spacer}}
\newcommand{\eldarica}{\textsc{Eldarica}}

\newcommand{\PP}{\ensuremath{\mathsf{P}}}
\newcommand{\A}{\ensuremath{\mathsf{A}}}
\newcommand{\LL}{\ensuremath{\mathsf{L}}}
\newcommand{\Tile}{\ensuremath{\mathsf{Tile}}}

\newcommand{\Inv}{\ensuremath{\mathsf{Inv}}}
\newcommand{\Ind}[1]{\ensuremath{\mathsf{Indices}}_{#1}}

\newcommand{\LB}{\ensuremath{\mathsf{L}_{\mathrm{body}}}}
\newcommand{\EE}{\ensuremath{\mathsf{E}}}
\newcommand{\PB}{\ensuremath{\mathsf{PB}}}
\newcommand{\Stmt}{\ensuremath{\mathsf{St}}}
\newcommand{\scVar}{\ensuremath{v}}
\newcommand{\lpVar}{\ensuremath{\ell}}
\newcommand{\ArVar}{\ensuremath{A}}
\newcommand{\BoolE}{\ensuremath{\mathsf{BoolE}}}
\newcommand{\iif}{\ensuremath{\mathbf{if}}}
\newcommand{\eelse}{\ensuremath{\mathbf{else}}}
\newcommand{\tthen}{\ensuremath{\mathbf{then}}}
\newcommand{\ffor}{\ensuremath{\mathbf{for}}}
\newcommand{\aassume}{\ensuremath{\mathbf{assume}}}
\newcommand{\cconst}{\ensuremath{\mathsf{c}}}
\newcommand{\PreC}{\ensuremath{\mathsf{PreCond}}}
\newcommand{\PostC}{\ensuremath{\mathsf{PostCond}}}
\newcommand{\AtomStmts}{\ensuremath{\mathsf{AtomSt}}}

\newcommand{\Unlabeled}{\ensuremath{\mathsf{U}}}
\newcommand{\RdAcc}[2]{\ensuremath{\mathsf{RdAcc}_{{#1}}({#2})}}
\newcommand{\CutPoints}{\ensuremath{\mathsf{CutPts}}}
\newcommand{\OuterVars}[1]{\ensuremath{\mathsf{OuterLoopCtrs}_{{#1}}}}
\newcommand{\UpdIndexExprs}[2]{\ensuremath{\mathsf{UpdIndexExprs}^{{#1}}[{#2}]}}
\newcommand{\InitTile}{\ensuremath{\mathsf{InitTile}}}
\newcommand{\infocomment}[1]{{\scriptsize\ttfamily\textcolor{darkgray}{\newline$\triangleright$ #1\newline}}}

\begin{document}

\title{Verifying Array Manipulating Programs by Tiling}

\author{Supratik Chakraborty$^1$ \and Ashutosh Gupta$^2$ \and Divyesh Unadkat$^{1,3}$}
\institute{Indian Institute of Technology Bombay, Mumbai, India\\
  \email{supratik@cse.iitb.ac.in} \and
  Tata Institute of Fundamental Research, Mumbai, India\\
  \email{agupta@tifr.res.in}
  \and
  TCS Research, Pune, India\\
  \email{divyesh.unadkat@tcs.com}}

\maketitle
\begin{abstract}
Formally verifying properties of programs that manipulate arrays in
loops is computationally challenging.  In this paper, we focus on a
useful class of such programs, and present a novel property-driven
verification method that first infers array access patterns in loops
using simple heuristics, and then uses this information to
compositionally prove universally quantified assertions about
arrays. Specifically, we identify \emph{tiles} of array accesses
patterns in a loop, and use the tiling information to reduce the
problem of checking a quantified assertion at the end of a loop to an
inductive argument that checks only a slice of the assertion for a
single iteration of the loop body.  We show that this method can be
extended to programs with sequentially composed loops and nested loops
as well. We have implemented our method in a tool called
$\ourtool$. Initial experiments show that $\ourtool$ outperforms
several state-of-the-art tools on a suite of interesting benchmarks.



\end{abstract}

\section{Introduction}
\label{sec:intro}
Arrays are widely used in programs written in imperative languages.
They are typically used to store large amounts of data in a region of
memory that the programmer views as contiguous, and which she can
access randomly by specifying an index (or offset).  Sequential
programs that process data stored in arrays commonly use looping
constructs to iterate over the range of array indices of interest and
access the corresponding array elements.  The ease with which data can
be accessed by specifying an index is often exploited by programmers
to access or modify array elements at indices that change in complex
ways within a loop.  While this renders programming easier, it also
makes automatic reasoning about such array manipulating programs
significantly harder.  Specifically, the pattern of array accesses
within loops can vary widely from program to program, and may not be
easy to predict.  Furthermore, since the access patterns often
span large regions of the array that depend on program parameters, the
array indices of interest cannot be bounded by statically estimated
small constants.  Hence, reasoning about arrays by treating each array
element as a scalar is not a practical option for analyzing such
programs.  This motivates us to ask if we can automatically infer
program-dependent patterns of array accesses within loops, and use
these patterns to simplify automatic verification of programs that
manipulate arrays in loops.

A commonly used approach for proving properties of sequential programs
with loops is to construct an inductive argument with an appropriate
loop invariant.  This involves three key steps: (i) showing that the
invariant holds before entering the loop for the first time, (ii)
establishing that if the invariant holds before entering the loop at
any time, then it continues to hold after one more iteration of the
loop, and (iii) proving that the invariant implies the desired
property when the loop terminates.  Steps (i) and (ii) allow us to
inductively infer that the invariant holds before every iteration of
the loop; the addition of step (iii) suffices to show that the desired
property holds after the loop terminates. A significant body of
research in automated program verification is concerned with finding
invariants that allow the above inductive argument to be applied
efficiently for various classes of programs.

For programs with loops manipulating arrays, the property of interest
at the end of a loop is often a universally quantified statement over
array elements.  Examples of such properties include $\forall i\,
\left((0 \le i < N) \rightarrow (A[i] \geq minVal) \wedge (A[i] \le
A[i+1])\right)$, $\forall i\, \left((0 \le i < N) \wedge (i \mod 2 = 0)
\rightarrow (A[i] = i)\right)$ and the like.  In such cases, a single
iteration of the loop typically only ensures that the desired property
holds over a small part of the array.  Effectively, each loop
iteration incrementally contributes to the overall property, and the
contributions of successive loop iterations compose to establish the
universally quantified property.  This suggests the following approach
for proving universally quantified assertions about arrays.
\begin{itemize}
\item We first \emph{identify the region of the array where the
  contribution of a generic loop iteration is localized}.  Informally,
  we call such a region a \emph{tile} of the array.  Note that
  depending on the program, the set of array indices representing a
  tile may not include all indices updated in the corresponding loop
  iteration.  Identifying the right tile for a given loop can be
  challenging in general; we discuss more about this later.
\item Next, we \emph{carve out a ``slice'' of the quantified property
  that is relevant to the tile identified above.}  Informally, we want
  this slice to represent the contribution of a generic loop iteration
  to the overall property.  The inductive step of our approach checks
  if a generic iteration of the loop indeed ensures this slice of the
  property.
\item Finally, we check that \emph{the tiles cover the entire range of
  array indices of interest, and successive loop iterations do not
  interfere with each other's contributions}.  In other words, once a
  loop iteration ensures that the slice of the property corresponding
  to its tile holds, subsequent loop iterations must not nullify this
  slice of the property.  Formalizing these ``range covering'' and
  ``non-interference'' properties allows us to show that the
  contributions of different loop iterations compose to yield the
  overall quantified property at the end of the loop.
\end{itemize} 
The remainder of the paper describes a technique and a tool that uses
the above ideas to prove quantified assertions in a useful class of
array manipulating programs.  We focus on assertions expressed as
universally quantified formulas on arrays, where the quantification is
over array indices.  Specifically, suppose $I$ denotes a sequence of
integer-valued array index variables, $A$ denotes an array and $\mathcal{V}$ denotes
a sequence of scalar variables used in the program.  We consider
assertions of the form $\forall I\, (\Phi(I) \implies \Psi(A,
\mathcal{V}, I))$, where $\Phi(I)$ is a quantifier-free formula in the
theory of arithmetic over integers, and $\Psi(A, \mathcal{V}, I)$ is a
quantifier-free formula in the combined theory of arrays and
arithmetic over integers.  Informally, such an assertion states that
for array indices satisfying condition $\Phi(I)$ (viz. even indices or
indices greater than a parameter $N$), the corresponding array
elements satisfy the property $\Psi(A, \mathcal{V}, I)$.  The formal
syntax of our assertions is explained in Section~\ref{sec:prelims}.
In our experience, assertions of this form suffice to express a large
class of interesting properties of array manipulating programs.

Although the general problem of identifying tiles in programs with
array manipulating loops is hard, we have developed some heuristics to
automate tile identification in a useful class of
programs. To understand the generic idea behind our tiling heuristic,
suppose the program under consideration has a single loop, and suppose
the quantified property is asserted at the end of the loop.  We
introduce a fresh counter variable that is incremented in each loop
iteration.  We then use existing arithmetic invariant generation
techniques, viz. \cite{invgen,daikon}, to identify a relation between
the indices of array elements that are accessed and/or updated in a
loop iteration, and the corresponding value of the loop counter.  This information
is eventually used to define a tile of the array for the loop under
consideration.

In a more general scenario, the program under verification may have a
sequence of loops, and the quantified property may be asserted at the
end of the last loop.  In such cases, we introduce a fresh counter
variable for each loop, and repeat the above process to identify a
tile corresponding to each loop.  For our tiling-based technique to
work, we also need invariants, or \emph{mid-conditions}, between
successive loops in the program.  Since identifying precise invariants
is uncomputable in general, we work with \emph{candidate invariants}
reported by existing off-the-shelf annotation/candidate-invariant
generators.  Specifically, in our implementation, we use the dynamic
analysis tool {\daikon}~\cite{daikon} that informs us of
\emph{candidate invariants} that are likely (but not proven) to hold
between loops.  Our algorithm then checks to see if the candidate
invariants reported after every loop can indeed be proved using the
tiling-based technique.  Only those candidates that can be proved in
this way are subsequently used to compose the tiling-based reasoning
across consecutive loops.  Finally, tiling can be applied to programs
with nested loops as well.  While the basic heuristic for identifying
tiles remains the same in this case, the inductive argument needs to
be carefully constructed when reasoning about nested loops.  We
discuss this in detail later in the paper.

We have implemented the above technique in a tool called {\ourtool}.
Our tool takes as input a C function with one or more loops
manipulating arrays.  It also accepts a universally quantified
assertion about arrays at the end of the function.  {\ourtool}
automatically generates a tiling of the arrays for each loop in the C
function and tries to prove the assertion, as described above.  We
have applied {\ourtool} to a suite of $60$ benchmarks comprised of
programs that manipulate arrays in different ways.  For most benchmarks
where the specified assertion holds, {\ourtool} was able to prove the
assertion reasonably quickly.  In contrast, two state-of-the-art tools
for reasoning about arrays faced difficulties and timed out on most of
these benchmarks.  For benchmarks where the specified assertion does
not hold, {\ourtool} relies on bounded model checking to determine if
an assertion violation can be detected within a few unwindings of the
loops.  There are of course corner cases where {\ourtool} remains
inconclusive about the satisfaction of the assertion.  Overall, our
initial experiments suggest tiling-based compositional reasoning can
be very effective for proving assertions in a useful class of array
manipulating programs.

The primary contributions of the paper can be summarized as follows.
\begin{itemize}
\item We introduce the concept of {\em tiling} for reasoning about
  quantified assertions in programs manipulating arrays in loops.
\item We present a tiling-based practical algorithm for verifying a
  class of array manipulating programs.
\item We describe a tool that outperforms several state-of-the-art
  tools for reasoning about arrays on a suite of benchmarks.  Our tool
  performs particularly well on benchmarks where the quantified
  assertion holds.
\end{itemize}



\section{Motivating Example}
\label{sec:motivating}
\begin{figure}[t]
  \begin{minipage}[b]{0.02\linewidth}
    \mbox{}
  \end{minipage}
\begin{minipage}[b]{0.4\textwidth}
\centering
\scriptsize
\lstset{language=C,
          numbers=left,
          numbersep=2pt,
          basicstyle=\ttfamily,
          keywordstyle=\color{black}\ttfamily,
          stringstyle=\color{red}\ttfamily,
          commentstyle=\color{green}\ttfamily,
          morecomment=[l][\color{magenta}]{\#}}
\begin{lstlisting}
void BatteryController() {
 int COUNT,MIN,i;
 int volArray[COUNT];

 if(COUNT%4 != 0) return;
 
 for(i=1;i<=COUNT/4;i++) {

  if(5 >= MIN)
   volArray[i*4-4] = 5;
  else
   volArray[i*4-4] = 0;  
  if(7 >= MIN)
   volArray[i*4-3] = 7;
  else
   volArray[i*4-3] = 0;
  if(3 >= MIN)
   volArray[i*4-2] = 3;
  else
   volArray[i*4-2] = 0;
  if(1 >= MIN)
   volArray[i*4-1] = 1;
  else
   volArray[i*4-1] = 0;  
 }
}
\end{lstlisting}
\vspace{-1ex}
\hspace{-10ex}
$\forall \mathtt{j}. ( 0 \leq \mathtt{j} < \mathtt{COUNT} \limplies \mathtt{volArray}[j]\geq\mathtt{MIN} 
\lor \mathtt{volArray}[j] =0)$\\
(a)
\end{minipage}
\begin{minipage}[b]{0.58\textwidth}
\centering
\scriptsize
\lstset{language=C,
          numbersep=2pt,
          numbers=left,
          basicstyle=\ttfamily,
          keywordstyle=\color{black}\ttfamily,
          stringstyle=\color{red}\ttfamily,
          commentstyle=\color{green}\ttfamily,
          morecomment=[l][\color{magenta}]{\#}}
\begin{lstlisting}
void BatteryControllerInst() {
  int COUNT,MIN,i,j;
  int volArray[COUNT];

  if(COUNT%4 != 0) return;

  assume(i>=1 && i<=COUNT/4);
  assume(4*i-4<=j && j<4*i);

  if(5 >= MIN) 
    volArray[i*4-4] = 5;
  else
    volArray[i*4-4] = 0;  
  if(7 >= MIN)
    volArray[i*4-3] = 7;
  else
    volArray[i*4-3] = 0;
  if(3 >= MIN)
    volArray[i*4-2] = 3;
  else
    volArray[i*4-2] = 0;
  if(1 >= MIN)
    volArray[i*4-1] = 1;
  else
    volArray[i*4-1] = 0;  
  assert(volArray[j]>=MIN||volArray[j]==0);
}
\end{lstlisting}
\vspace{1.5ex}
(b)
\end{minipage}
\caption{Motivating example \texttt{period-4}}
\label{fig:motex}
\end{figure}

Fig.~\ref{fig:motex}(a) shows a C function snippet adapted from an
industrial battery controller.  This example came to our attention
after a proprietary industry-strength static analysis tool failed to
prove the quantified assertion at the end of the function.  Note that
the function updates an array {\tt volArray} whose size is given by
{\tt COUNT}.  In general, {\tt COUNT} can be large, viz.  $100000$.
The universally quantified assertion at the end of the ``for'' loop
requires that every element of {\tt volArray} be either zero or at
least as large as {\tt MIN}.  It is not hard to convince oneself
through informal reasoning that the assertion indeed holds.  The
difficulty lies in proving it automatically. Indeed, neither
{\booster}~\cite{booster} nor {\vaphor}~\cite{vaphor}, which can
reason about arrays with parameterized bounds, are able to prove this
assertion within 15 minutes on a desktop machine. Bounded model
checking tools like {\cbmc}~\cite{cbmc} and
{\smackpluscorral}~\cite{corral} are able to prove this assertion for
arrays with small values of {\tt COUNT}.  For large arrays, viz. {\tt
  COUNT = 100000}, these tools cannot prove the assertion within 15
minutes on a desktop machine.  This is not surprising since a bounded
model checker must unwind the loop in the function a large number of
times if {\tt COUNT} is large. 

Let us now illustrate how tiling-based reasoning works in this
example.  We introduce a fresh auxiliary variable (say {\tt j}) to
denote the index used to update an element of {\tt
  volArray}.  Using arithmetic invariant generation techniques,
viz.{\invgen}~\cite{invgen}, we can now learn that for all array accesses
in the $i^{th}$ loop iteration, the value of the index lies between
$\mathtt{4*i-4}$ and $\mathtt{4*i}$.  Therefore, we choose
$[\mathtt{4*i-4},\mathtt{4*i})$ as the tile corresponding to the
  $i^{th}$ iteration of the loop.  

  In order to successfully apply the tiling-based reasoning, we must
  ensure that our tiles satisfy certain properties.
\begin{itemize}
\item \emph{Covers range}: This ensures that every tile contains only
  valid array indices, and that no array index of interest in the
  quantified assertion is left unaccounted for in the tiles.  In our
  example, array indices range
  from $0$ to $\mathtt{COUNT}-1$, while the loop (and hence, tile) counter
  $\mathtt{i}$ ranges from $1$ to $\mathtt{COUNT}/4$.
  Since the $i^{th}$ tile comprises of the array indices $4i-4, 4i-3, 4i-2$ and
  $4i-1$, both the above requirements are met.

\item \emph{Sliced property holds for tile}: The sliced property in
  this case says that the elements of {\tt volArray} corresponding to
  indices within a tile have values that are either $0$ or at least
  {\tt MIN}.  To prove that this holds after an iteration of the loop,
  we first obtain a loop-free program containing a single generic
  iteration of the loop, and check that the elements of {\tt volArray}
  corresponding to the $i^{th}$ tile satisfy the sliced property after
  the execution of the $i^{th}$ loop iteration.  The transformed
  program is shown in Fig.~\ref{fig:motex}(b).  Note that this program
  has a fresh variable $j$.  The assume statements at lines $7$-$8$
  say that $i$ is within the expected range and that $j$ is an index
  in the $i^{th}$ tile.  Since this program is loop-free, we can use a
  bounded model checker like {\cbmc}~\cite{cbmc} to prove the
  assertion in the transformed program.

\item \emph{Non-interference across tiles}: To show this, we assume
  that the sliced property holds for the $i'$-th tile, where $0 \le i' < i$,
  before the $i^{th}$ loop iteration starts.  This can be done
  by adding the following three \emph{extra assumptions} after lines
  $7$ and $8$ in Fig.~\ref{fig:motex}(b): (i) {\tt assume (1 <= i' <
    i)}, (ii) {\tt assume (4*i' - 4 <= j' < 4*i')}, and (iii) {\tt
    assume (volArray[j'] >= MIN || volArray[j'] == 0)}. We then assert
  at the end of the loop body that the sliced property for the $i'$-th
  tile continues to hold even after the $i^{th}$ iteration.  This can
  be done by replacing the assertion in line $26$ of
  Fig.~\ref{fig:motex}(b) by {\tt assert (volArray[j'] >= MIN ||
    volArray[j'] == 0)}.  As before, since the program in
  Fig.~\ref{fig:motex}(b) is loop-free, this assertion can be easily
  checked using a bounded model checker like {\cbmc}.
\end{itemize}
Once all the above checks have succeeded, we can conclude that the
quantified assertion holds in the original program after the loop
terminates.  Note the careful orchestration of inductive reasoning to
prove the sliced property, and compositional reasoning to aggregate
the slices of the property to give the original quantified assertion.
Our tiling-based tool proves the assertion in this example in less
than a second.



\section{Preliminaries}
\label{sec:prelims}
For purposes of this paper, an array-manipulating program $\PP$ is a
tuple $(\mathcal{V}, \mathcal{L}, \mathcal{A}, {\PB})$, where
$\mathcal{V}$ is a set of scalar variables, $\mathcal{L} \subseteq
\mathcal{V}$ is a set of scalar loop counter variables, $\mathcal{A}$
is a set of array variables, and ${\PB}$ is the program body generated
by the following grammar.
\begin{center}
\begin{tabular}{rcl}
  {\PB} & ::= & \Stmt\\
  \Stmt & ::= & {\scVar} := \EE ~$\mid$~ {\ArVar}[\EE] := \EE ~$\mid$~
                {\aassume}(\BoolE) ~$\mid$~
                {\iif}(\BoolE) {\tthen} {\Stmt} {\eelse} \Stmt ~$\mid$~\\
        &     & {\ffor} ({\lpVar} := 0; {\lpVar} $<$ \EE; {\lpVar} := {\lpVar}+1) ~\{{\Stmt}\}~ ~$\mid$~
                \Stmt~;~\Stmt \\
 \EE & ::=  &\EE ~op~ \EE ~$\mid$~ {\ArVar}[\EE] ~$\mid$~ {\scVar} ~$\mid$~ {\lpVar} ~$\mid$~ {\cconst} \\
 \BoolE & ::= & \EE ~relop \EE ~$\mid$~ {\BoolE} AND {\BoolE} ~$\mid$~ NOT {\BoolE} ~$\mid$~ {\BoolE} OR {\BoolE}
\end{tabular}
\end{center}
\noindent Here, we assume that ${\ArVar} \in \mathcal{A}$, ${\scVar} \in
\mathcal{V}\setminus \mathcal{L}$, ${\lpVar} \in \mathcal{L}$ and
${\cconst}\in \integers$.  We also assume that ``op''
(resp. ``relop'') is one of a set of arithmetic (resp. relational)
operators.  We wish to highlight the following features of programs
generated by this grammar:
\begin{itemize}
\item There are no unstructured jumps, like those effected by {\tt
  goto} or {\tt break} statements in C-like languages.  The effect of
  a {\tt break} statement inside a loop in a C-like language can
  always be modeled by setting a flag, and by conditioning the
  execution of subsequent statements in the loop body on this flag
  being not set, and by using this flag to determine whether to exit
  the loop.  The effect of a {\tt break} statement in a conditional
  branch can also be similarly modeled.  Therefore, we can mimic the
  behaviour of {\tt break} statements in our programs.
\item We can have sequences of possibly nested loops, with non-looping
  program fragments between loops. Furthermore, the body of a loop and
  the corresponding \emph{loop head}, i.e. control location where the
  loop is entered, are easily identifiable.
\item Every loop is associated with a scalar loop counter variable
  that is set to $0$ when the loop is entered, and incremented after
  every iteration of the loop.  We assume that each loop has a unique
  counter variable.
\item The only assignments to loop counter variables happen when a
  loop is entered for the first time and at the end of an iteration of
  the corresponding loop body.  Other assignment statements in the
  program cannot assign to loop counter variables.  Loop counter
  variables can however be freely used in expressions throughout the
  program.
\item The restriction on the usage of loop counter variables
  simplifies the analysis and presentation, while still allowing a
  large class of programs to be effectively analyzed.  Specifically,
  whenever the count of iterations of a loop can be expressed in a
  closed form in terms of constants and variables not updated in the
  loop, we can mimic its behaviour using our restricted loops.  As a
  generic example, suppose we are told that the loop {\tt for
  (i:=exp1; Cond; i:=exp2) \{ LoopBody \}} iterates {\tt exp3} times,
  where {\tt exp3} is an arithmetic expression in terms of constants
  and variables not updated in the loop.  The behaviour of this loop
  can be mimicked using the following restricted loop, where {\tt l}
  and {\tt flag} are fresh variables not present in the original
  program: {\tt for (l:=0; l<exp3 ; l:=l+1) \{if (l=0) \{i:=exp1\};
  if (Cond) \{ LoopBody; i:=exp2\}\}}.

  To see a specific example of this transformation, suppose the
  program under verification has the loop: {\tt for (i:=2*M; i>=0;
  i:=i-2)\{ LoopBody \}}, where {\tt M} and {\tt i} are variables not
  updated in {\tt LoopBody}.  Clearly, this loop iterates $(M+1)$
  times.  Therefore, it can be modeled in our restricted language as:
  {\tt for (l:=0; l<M+1; l:=l+1) \{ if (l=0) \{i:=2*M\}; if (i >=
  0) \{ LoopBody; i:=i-2\}\}}.
\end{itemize}

For clarity of exposition, we abuse notation and use ${\mathcal{V}}$
and ${\mathcal{A}}$ to also denote a sequence of scalar and array
variables, when there is no confusion. A verification problem for an
array manipulating program is a Hoare triple $\{\PreC\} ~{\PP}~
\{\PostC\}$, where each of ${\PreC}$ and ${\PostC}$ are quantified
formulae of the form $\forall I\, \left(\Phi(I) \implies
\Psi(\mathcal{A},\mathcal{V}, I)\right)$.  Here, $I$ is assumed to be
a sequence of array index variables, $\Phi$ is a quantifier-free
formula in the theory of arithmetic over integers, and $\Psi$ is a
quantifier-free formula in the combined theory of arrays and
arithmetic over integers.  The formula $\Phi(I)$ identifies the
relevant indices of the array where the property $\Psi(\mathcal{A},
\mathcal{V}, I)$ must hold.  This allows us to express a large class
of useful pre- and post-conditions, including sortedness, which can be
expressed as $\forall j\, (0 \le j < N) \rightarrow (A[j] \le A[j+1])$.

Let {\AtomStmts} denote the set of atomic statements in a program
generated by the above grammar. These are statements of the form
${\scVar} := \EE$, ${\ArVar}[\EE] := \EE$ or ${\aassume}(\EE)$.
It is common to represent such a program by a \emph{control flow
graph} $G = (N, E, \mu)$, where $N$ denotes the set of control
locations of the program, $E \subseteq N \times N \times \{\ltrue,
\lfalse, \Unlabeled\}$ represents the flow of control, and $\mu: N
\rightarrow {\AtomStmts} \cup {\BoolE}$ annotates every node in $N$
with either an assignment statement, an assume statement or a Boolean
condition.
\begin{wrapfigure}[19]{r}{0.3\textwidth}
\vspace*{-0.1in}
\begin{center}
{\scriptsize
  \begin{tikzpicture}[%
    ->,
    shorten >=2pt,
    >=stealth,
    node distance=0.4cm,
    noname/.style={%
      ellipse,
      minimum width=1em,
      minimum height=0.5em,
      draw
    }
  ]
    \node[noname] (0)              {S};
    \node[noname] (1) [below=of 0] {1};
    \node[noname] (2) [below=of 1] {2};
    \node[noname] (3) [below=of 2] {3};
    \node[noname] (4) [below left=of 3] {4};
    \node[noname] (6) [below right=of 3] {6};
    \node[noname] (5) [below right=of 4] {5};
    \node[noname] (7) [below=of 5] {7};
    \node[noname] (8) [below=of 7] {E};

    \path (0) edge                   node {} (1)
          (1) edge                   node {$~~~~\ltrue$} (2)
          (2) edge                   node {$\ltrue~~~~$} (3)
          (3) edge                   node {$\ltrue~~~~$} (4)
          (3) edge                   node {$~~~~\lfalse$} (6)
          (4) edge                   node {} (5)
          (6) edge                   node {$~~~~\ltrue$} (5)
          (5) edge                   node {$~~~e_2$} (3)
          (6) edge [bend right=69pt] node {$e_1~\lfalse$} (2)
          (7) edge [bend right=99pt] node {$~~~~e_3$} (1)
          (2) edge [bend right=99pt] node {$\lfalse~~~$} (7)
          (1) edge [bend right=86pt] node {$\lfalse~~~$} (8);
  \end{tikzpicture}
}
\end{center}
\caption{A CFG}
\label{fig:cfg}
\end{wrapfigure}


We assume there are two distinguished vertices called
$\mathsf{Start}$ and $\mathsf{End}$ in $N$, that represent the entry
and exit points of control flow for the program.  An edge $(n_1, n_2,
L)$ represents flow of control from $n_1$ to $n_2$ without any other
intervening node.  The edge is labeled $\ltrue$ or $\lfalse$ if
$\mu(n_1)$ is a Boolean condition, and it is labeled $\Unlabeled$
otherwise.  If $\mu(n_1)$ is a Boolean condition, there are two
outgoing edges labeled $\ltrue$ and $\lfalse$ respectively, from
$n_1$.  Control flows from $n_1$ to $n_2$ along $(n_1, n_2, L)$ only
if $\mu(n_1)$ evaluates to $L$.  If $\mu(n_1)$ is an assume or
assignment statement, there is a single outgoing edge from $n_1$, and
it is labeled $\Unlabeled$.  Henceforth, we use CFG to refer to a
control flow graph.  

A CFG may have cycles in general.  A \emph{back-edge} in a CFG is an
edge from a node (control location) within the body of a loop to the
node representing the corresponding loop head.  Clearly, removing all
back-edges from a CFG renders it acyclic.  The target nodes of
back-edges, i.e. nodes corresponding to loop heads, are also
called \emph{cut-points} of the CFG.  Every acyclic sub-graph of a CFG
that starts from a cut-point or $\mathsf{Start}$ and ends at another
cut-point or $\mathsf{End}$, and that does not pass through any
other cut-points in between and also does not include any back-edge,
is called a \emph{segment}.  For example, consider the CFG shown in
Fig.~\ref{fig:cfg}.  For clarity, edges labeled $\Unlabeled$ are
shown unlabeled in the figure.  The cut-points in this CFG are nodes
$1$, $2$ and $3$, the back-edges are $e_1$, $e_2$ and $e_3$, and the
segments are $S\rightarrow 1$, $1 \rightarrow 2$, $2 \rightarrow 3$,
$3 \rightarrow \{4, 6\} \rightarrow 5$, $2\rightarrow 7$ and
$1\rightarrow E$. Note that every segment is an acyclic sub-graph of
the CFG with a unique source node and a unique sink node.

\section{A Theory of Tiles}
\label{sec:tiling}
In this section, we present a theory of tiles for proving universally
quantified properties of arrays in programs that manipulate arrays
within loops. 
\subsection{Tiling in a simple setting}
\label{sec:tiling-simple}
Consider a program {\PP} as defined in the previous section that
accesses elements of an array {\A} in a loop {\LL}.  Suppose {\PP} has
a single non-nested loop {\LL} with loop counter $\ell$ and loop exit
condition $\mexit{\ell}{\ell}$, where $\mE_{\ell}$ is an arithmetic
expression involving only constants and variables not updated in
{\LL}.  Thus, the loop iterates ${\mE}_{\ell}$ times, with the value
of $\ell$ initialized to $0$ at the beginning of the first iteration,
and incremented at the end of each iteration.  Each access of an
element of {\A} in the loop is either a \emph{read access} or a
\emph{write access}.  For example, in the program shown in
Fig.~\ref{fig:simplecode}, the loop {\LL} (lines $2$-$11$) has three
read accesses of {\A} (at lines $5$, $6$, $7$), and three write
accesses of {\A} (at lines $3$, $6$, $7$).  In order to check an
assertion about the array at the end of the loop (see, for example,
line $12$ of Fig.~\ref{fig:simplecode}), we wish to tile the array
based on how its elements are updated in different iterations of the
loop, reason about the effect of each loop iteration on the
corresponding tile, and then compose the tile-wise reasoning to
prove/disprove the overall assertion.
\begin{wrapfigure}[10]{R}{0.53\textwidth}
\vspace*{-0.4in}
{\scriptsize
\begin{verbatim}
1. void ArrayUpdate(int A[], int n) {
2.   for (int l:=0; l < n; l:=l+1) { // loop L
3.     if ((l = 0) OR (l = n-1)) {A[l] := THRESH;}
4.     else {
5.       if (A[l] < THRESH) { 
6.         A[l+1] := A[l] + 1;
7.         A[l] := A[l-1];
8.       } // end if
9.     } // end else
11.  } // end for
12.  // assert(forall i in 0..n-1, A[i] >= THRESH);
13. }
\end{verbatim}
}
\caption{Program with interesting tiling}\label{fig:simplecode}
\end{wrapfigure}
Note that the idea of tiling an
array based on access patterns in a loop is not new, and has been used
earlier in the context of parallelizing and optimizing
compilers~\cite{SSK17,TreeTiler}.  However, its use in the context of
verification has been limited~\cite{ArrayCousotCL11}.  To explore the
idea better, we need to formalize the notion of \emph{tiles}.

Let $\Ind{A}$ denote the range of indices of the array {\A}.  We
assume that this is available to us; in practice, this can be obtained
from the declaration of {\A} if it is statically declared, or from the
statement that dynamically allocates the array {\A}.  Let $\ppre$ and
$\ppost$ denote the pre- and post-conditions, respectively, for the
loop {\LL} under consideration.  Recall from Section~\ref{sec:prelims}
that both $\ppre$ and $\ppost$ have the form $\forall j\,
\left(\Phi(j) \implies \Psi(\A, \mathcal{V}, j)\right)$, where
$\mathcal{V}$ denotes the set of scalar variables in the program.  To
keep the discussion simple, we consider $\ppost$ to be of this
specific form for the time being, while ignoring the form of $\ppre$.
We show later how the specific form of $\ppre$ can be used to simplify
the analysis further.  For purposes of simplicity, we also assume that
the array {\A} is one-dimensional; our ideas generalize easily to
multi-dimensional arrays, as shown later.  Let $\Inv$ be a (possibly
weak) loop invariant for loop {\LL}.  Clearly, if $\ppre \implies
\Inv$ and $\Inv \wedge \neg\mexit{\ell}{\LL} \implies \ppost$, then we
are already done, and no tiling is necessary.  The situation becomes
interesting when $\Inv$ is not strong enough to ensure that $\Inv
\wedge \neg\mexit{\ell}{\LL} \implies \ppost$.  We encounter several
such cases in our benchmark suite, and it is here that our method adds
value to existing verification flows.

A \emph{tiling} of {\A} with respect to {\LL}, $\Inv$ and $\ppost$ is
a binary predicate $\Tile_{\LL, \Inv, \ppost}: \mathbb{N} \times
\Ind{\A} \rightarrow \{\ltrue,\lfalse\}$ such that conditions T1
through T3 listed below hold.  Note that these conditions were
discussed informally in Section~\ref{sec:motivating} in the context of
our motivating example.  For ease of notation,
we use $\Tile$ instead of $\Tile_{\LL, \Inv, \ppost}$ below, when
${\LL}$, $\Inv$ and $\ppost$ are clear from the context.  We also use
``$\ell^{th}$ tile'' to refer to all array indices in the set $\{j
\mid (j \in \Ind{\A}) \wedge \Tile(\ell, j)\}$.
\begin{itemize}
\item[(T1)] \emph{Covers range:}
Every array index of interest must be present in some tile, and every
tile contains array indices in $\Ind{\A}$.  Thus, the formula
$\eta_1 \wedge \eta_2$ must be valid, where $\eta_1 ~\equiv~ \forall
j \left((j \in \Ind{\A}) \wedge \Phi(j) \implies \exists
\ell \, \left((0 \le \ell < {\mE}_{\ell})  \wedge \Tile(\ell, j)\right)\right)$,
and $\eta_2 ~\equiv~ \forall \ell\, \left((0 \le \ell <
{\mE}_{\ell}) \wedge \Tile(\ell, j) \implies (j \in \Ind{\A})\right)$.

\item[(T2)] \emph{Sliced post-condition holds inductively:}
We define the sliced post-condition for the $\ell^{th}$ tile as
$\ppost_{\Tile(\ell,\cdot)} \defeq \forall j \, (\Tile(\ell,
j) \wedge \Phi(j) \implies \Psi(\A, \mathcal{V}, j))$.  Thus,
$\ppost_{\Tile(\ell,\cdot)}$ asserts that $\Psi(\A, \mathcal{V}, j)$
holds for all relevant $j$ in the $\ell^{th}$ tile.  We now require that if the
(possibly weak) loop invariant ${\Inv}$ and the sliced post-condition
for the $\ell'$-th tile for all $\ell' \in \{0, \ldots \ell-1\}$ hold
prior to executing the $\ell^{th}$ loop iteration, then the sliced
post condition for the $\ell$-th tile and ${\Inv}$ must also hold after
executing the $\ell^{th}$ loop iteration.

Formally, if {\LB} denotes the body of the loop {\LL}, the Hoare
triple given by $\{\Inv \wedge \bigwedge_{\ell': 0 \le \ell' < \ell}
\ppost_{\Tile(\ell', \cdot)}\} ~{\LB}~ \{\Inv \wedge \ppost_{\Tile(\ell, \cdot)}\}$
must be valid for all $\ell \in \{0, \ldots \mE_{\ell}-1\}$. 

\item[(T3)] \emph{Non-interference across tiles:}
For every pair of iterations $\ell, \ell'$ of the loop {\LL} such that $\ell'
< \ell$, the later iteration ($\ell$) must not falsify the sliced post
condition $\ppost_{\Tile(\ell',\cdot)}$ rendered true by the earlier
iteration ($\ell'$).

Formally, the Hoare triple $\{\Inv \wedge (0 \le \ell'
< \ell) \wedge \ppost_{\Tile(\ell', \cdot)}\}
{\LB} \{\ppost_{\Tile(\ell', \cdot)}\}$ must be valid for all
$\ell \in \{0, \ldots \mE_{\ell}-1\}$.
\end{itemize}
Note that while tiling depends on $\LL$, $\Inv$ and $\ppost$ in
general, the pattern of array accesses in a loop often suggests a
natural tiling of array indices that suffices to prove multiple
assertions $\ppost$ using reasonably weak loop invariants $\Inv$.  The
motivating example in Section~\ref{sec:motivating} illustrated this
simplification.  The example in Fig.~\ref{fig:simplecode} admits the
tiling predicate $\Tile(\ell, j) \equiv (j = \ell)$ based on
inspection of array access patterns in the loop.  Note that in this
example, the $\ell^{th}$ iteration of the loop can update both
${\A}[\ell]$ and ${\A}[\ell+1]$.  However, as we show later, a simple
reasoning reveals that the right tiling choice here is $(j = \ell)$,
and not $(\ell \le j \le \ell+1)$.

\begin{theorem}\label{thm:sound}
Suppose $\Tile_{\LL, \Inv, \ppost}: \mathbb{N} \times \Ind{\A}
\rightarrow \{\ltrue,\lfalse\}$ satisfies conditions T1 through T3. If
$\ppre \implies \Inv$ also holds and the loop {\LL} iterates at least
once, then the Hoare triple $\{\ppre\} ~{\LL}~ \{\ppost\}$ holds.
\end{theorem}
\noindent \emph{Proof sketch:} The proof proceeds by induction on the
values of the loop counter $\ell$.  The inductive claim is that at the
end of the $\ell^{th}$ iteration of the loop, the post-condition
$\bigwedge_{\ell': 0 \le \ell' \leq \ell} \ppost_{\Tile(\ell',
  \cdot)}$ holds.  The base case is easily seen to be true from
condition T2 and from the fact that $\ppre \implies \Inv$.  Condition
T3 and the fact that $\ell$ is incremented at the end of each loop
iteration ensure that once we have proved $\ppost_{\Tile(\ell,
  \cdot)}$ at the end of the $\ell^{th}$ iteration, it cannot be
falsified in any subsequent iteration of the loop.  Condition T2 now
ensures that the sliced post-condition can be inductively proven for
the $\ell^{th}$ tile. By condition T1, we also have $\bigwedge_{0 \le
  \ell < {\mE_{\ell}}} \ppost_{\Tile(\ell,\cdot)} \equiv \ppost$.  Since
the loop {\LL} iterates with $\ell$ increasing from $0$ to
$\mE_{\ell}-1$, it follows that $\ppost$ indeed holds if {\Inv} holds
before the start of the first iteration. This is the compositional
step in our approach.  Putting all the parts together, we obtain a
proof of $\{\ppre\} ~{\LL}~ \{\ppost\}$. \qed

A few observations about the conditions are worth noting.  First, note
that there is an alternation of quantifiers in the check for T1.
Fortunately, state-of-the-art SMT solvers like {\zthree}~\cite{z3}
are powerful enough to check this condition efficiently for tiles
expressed as Boolean combinations of linear inequalities on $\ell$ and
$\mathcal{V}$, as is the case for the examples in our benchmark suite.
We anticipate that with further advances in reasoning about
quantifiers, the check for condition T1 will not be a
performance-limiting step.

The checks for T2 and T3 require proving Hoare triples with
post-conditions that have a conjunct of the form $\ppost_{\Tile(\ell,
  \cdot)}$.  From the definition of a sliced post-condition, we know
that $\ppost_{\Tile(\ell, \cdot)}$ is a universally quantified
formula.  Additionally, the pre-condition for T2 has a conjunct of the
form $\bigwedge_{\ell': 0 \le \ell' < \ell} \ppost_{\Tile(\ell',
  \cdot)}$, which is akin to a universally quantified formula.
Therefore T2 and T3 can be checked using Hoare logic-based reasoning
tools that permit quantified pre- and post-conditions,
viz.~\cite{key-hoare,verifast}.  Unfortunately, the degree
of automation and scalability available with such tools is limited
today.  To circumvent this problem, we propose to use stronger Hoare
triple checks that logically imply T2 and T3, but do not have
quantified formulas in their pre- and post-conditions.  Since the
program, and hence {\LB}, is assumed not to have nested loops,
state-of-the-art bounded model checking tools that work with
quantifier-free pre- and post-conditions, viz. {\cbmc}, can be used to
check these stronger conditions.  Specifically, we propose the
following pragmatic replacements of T2 and T3.
\begin{itemize}
\item[(T2*)] Let $\RdAcc{\LL}{\ell}$ denote the set of array index expressions
corresponding to read accesses of {\A} in the $\ell^{th}$ iteration of
the loop {\LL}.  For example, in Fig.~\ref{fig:simplecode},
$\RdAcc{\LL}{\ell} = \{\ell, \ell-1\}$.  Clearly, if {\LB} is
loop-free, $\RdAcc{\LL}{\ell}$ is a finite set of expressions.  Suppose
$|\RdAcc{\LL}{\ell}| = k$ and let $e_1, \ldots e_k$ denote the expressions
in $\RdAcc{\LL}{\ell}$.  Define $\zeta(\ell)$ to be the formula
$\bigwedge_{e_k\in \RdAcc{\LL}{\ell}} \big(\left((0 \le \ell_k < \ell
< \mE_{\ell}) \wedge \Tile(\ell_k, e_k) \wedge
\Phi(e_k)\right) \Rightarrow \Psi({\A}, \mathcal{V}, e_k)\big)$, where
$\ell_k$ are fresh variables not used in the program.  Informally,
$\zeta(\ell)$ states that if $\A[e_k]$ is read in the $\ell^{th}$
iteration of {\LL} and if $e_k$ belongs to the $\ell_k$-th ($\ell_k
< \ell$) tile, then $\Phi(e_k) \implies \Psi(\A, \mathcal{V}, e_k)$
holds.

We now require the following Hoare triple to be valid, where $j$ is
a fresh free variable not used in the program.\\
$\{\Inv \wedge (0 \le \ell < \mE_{\ell}) \wedge \zeta(\ell) \wedge \Tile(\ell, j) \wedge \Phi(j)\}
~{\LB} \{\Inv \wedge \Psi({\A}, \mathcal{V}, j)\}$.  
\item[(T3*)]
Let $j'$ and $\ell'$ be fresh free variables that are not used in the program.
We require the following Hoare triple to be valid:\\
$\{\Inv \wedge (0 \le \ell' < \ell < \mE_{\ell}) \wedge \Tile(\ell',
j') \wedge \Phi(j')
\wedge \Psi(\A, \mathcal{V}, j')\} ~{\LB}~ \{\Psi({\A}, \mathcal{V}, j')\}$ 
\end{itemize}
\begin{lemma}
\label{lem:simp-checks}
The Hoare triple in T2* implies that in T2.  Similarly, the Hoare triple
in T3* implies that in T3.
\end{lemma}
The proof follows from the observation that a counterexample for
validity of the Hoare triple in T2 or T3 can be used to construct a
counterexample for validity of the triple in T2* or T3* respectively.

Observe that T2* and T3* require checking Hoare triples with
quantifier-free formulas in the pre- and post-conditions.  This makes
it possible to use assertion checking tools that work with
quantifier-free formulas in pre- and post-conditions.  Furthermore,
since {\LB} is assumed to be loop-free, these checks can
also be discharged using state-of-the-art bounded model checkers,
viz. {\cbmc}.  The scalability and high degree of automation provided
by tools like {\cbmc} make conditions T1, T2* and T3* more attractive to use.

\subsection{Tiling in more general settings}
\label{sec:tiling-general}
The above discussion was restricted to a single uni-dimensional array
accessed within a single non-nested loop in a program {\PP}.  We now
relax these restrictions and show that the same technique
continues to work with some adaptations.

We consider the case where {\PP} is a sequential composition of
possibly nested loops.  To analyze such programs, we identify all
segments in the CFG of {\PP}.  Let {\CutPoints} be the set of
cut-points of the CFG.  Recall from Section~\ref{sec:prelims} that a
segment is a sub-DAG of the CFG between a source node in ${\CutPoints}
\cup \{\mathsf{Start}\}$ and a sink node in ${\CutPoints} \cup
\{\mathsf{End}\}$.  Thus, a segment $s$ corresponds to a loop-free
fragment of ${\PP}$.  Let $\ell_s$ denote the loop counter variable
corresponding to the innermost loop in which $s$ appears.  We assign
$\bot$ to $\ell_s$ if $s$ lies outside all loops in {\PP}.  Let
{\OuterVars{s}} denote the set of loop counter variables of all outer
loops (excluding the innermost one) that enclose (or nest) $s$.  The
syntactic restrictions of programs described in
Section~\ref{sec:prelims} ensure that $\ell_s$ and {\OuterVars{s}} are
uniquely defined for every segment $s$.

Suppose we are given (possibly weak) invariants at every cut-point in
${\PP}$, where ${\Inv}_c$ denotes the invariant at cut-point $c$.  We
assume the invariants are of the usual form $\forall I\, \left(\Phi(I)
\implies \Psi(\mathcal{A}, \mathcal{V}, I)\right)$, where $I$ is a
sequence of quantified array index variables, and $\mathcal{A}$ and
$\mathcal{V}$ are sequences of array and scalar variables
respectively.  Let $\mathcal{A}_s$ be a sequence of arrays that are
updated in the segment $s$ between cut-points $c_1$ and $c_2$, and for
which $\ell_s \neq \bot$.  We define a tiling predicate $\Tile_{s,
  {\Inv}_{c_1}, {\Inv}_{c_2}}: \mathbb{N} \times \Ind{\mathcal{A}_s}
\rightarrow \{\ltrue,\lfalse\}$, where $\Ind{\mathcal{A}_s} =
\prod_{\A' \in \mathcal{A}_s} \Ind{\A'}$ plays a role similar to that
of $\Ind{\A}$ in Section~\ref{sec:tiling-simple} (where a single array
{\A} was considered).  The predicate $\Tile_{s, {\Inv}_{c_1},
  {\Inv}_{c_2}}$ relates values of the loop counter $\ell_s$ of the
innermost loop containing $s$ to the index expressions that define the
updates of arrays in $\mathcal{A}_s$ in the program segment $s$.  The
entire analysis done in Section~\ref{sec:tiling-simple} for a simple
loop {\LL} can now be re-played for segment $s$, with ${\Inv}_{c_1}$
playing the role of ${\Inv}$, ${\Inv}_{c_2}$ playing the role of
$\ppost$, $\mathcal{V} \cup \OuterVars{s}$ playing the role of
$\mathcal{V}$, and $\ell_s$ playing the role of $\ell$.  If the
segment $s$ is not enclosed in any loop, i.e. $\ell_s = \bot$,
we need not define any tiling predicate for this segment.  This
obviates the need for conditions T1 and T3, and checking T2 simplifies
to checking the validity of the Hoare triple $\{{\Inv}_{c_1}\} ~s~
\{{\Inv}_{c_2}\}$.  In general, ${\Inv}_{c_1}$ and ${\Inv}_{c_2}$ may
be universally quantified formulas.  In such cases, the technique used
to simplify condition T2 to T2* in Section~\ref{sec:tiling-simple} can
be applied to obtain a stronger condition, say T2**, that does not
involve any tile, and requires checking a Hoare triple with
quantifier-free pre- and post-conditions. If the condition checks for
all segments as described above succeed, it follows from
Theorem~\ref{thm:sound} and Lemma~\ref{lem:simp-checks} that we have a
proof of $\{\ppre\} ~{\PP}~ \{\ppost\}$.

Recall that in Section~\ref{sec:tiling-simple}, we ignored the
specific form of the pre-condition $\ppre$.  As defined in
Section~\ref{sec:prelims}, $\ppre$ has the same form as that of the
post-condition and invariants at cut-points considered
above. Therefore, the above technique works if we treat $\ppre$ as
${\Inv}_{\mathsf{Start}}$ and $\ppost$ as ${\Inv}_{\mathsf{End}}$.

The extension to multi-dimensional arrays is straightforward.  Instead
of using one index variable $j$ for accessing arrays, we now allow a
tuple of index variables $(j_1, j_2, \ldots j_r)$ for accessing
arrays.  Each such variable $j_l$ takes values from its own domain,
say $\Ind{\A_l}$.  The entire discussion about tiles above continues
to hold, including the validity of Theorem~\ref{thm:sound}, if we
replace every occurrence of an array index variable $j$ by a sequence
of variables $j_1, \ldots j_r$ and every occurrence of $\Ind{\A}$ by
$\Ind{\A_1} \times \Ind{\A_2} \ldots \times \Ind{\A_r}$.  





\section{Verification by Tiling}
\label{sec:algos}
The discussion in the previous section suggests a three-phase
algorithm, presented as Algorithm~\ref{alg:overall}, for verifying
quantified properties of arrays in programs with sequences of possibly
nested loops manipulating arrays.  In the first phase of the
algorithm, we use bounded model checking with small pre-determined
loop unrollings to check for assertion violations.  If this fails, we
construct the CFG of the input program {\PP}, topologically sort its
cut-points and initialize the sets of candidate invariants at each
cut-point to $\emptyset$.

\begin{algorithm}[!th]
 \caption{\textsc{TiledVerify}({\PP} : program, $\ppre$: pre-condn, $\ppost$: post-condn)}
 \label{alg:overall}
\begin{algorithmic}[1]
 \scriptsize
 \State Let $G$ be the CFG for program
 ${\PP} = (\mathcal{A}, \mathcal{V}, \mathcal{L}, {\PB})$, as defined in
 Section~\ref{sec:prelims}.

 \infocomment{Check for shallow counterexample and initialization}
 \State Do bounded model checking with pre-determined small loop unrollings;
 \If{counterexample found}
 \Return{``Post condition violated!''};
 \EndIf
 \State {\CutPoints} := set of cut-points in $G$;
 \State Remove all back-edges from $G$ and topologically sort {\CutPoints};
        \Comment{Let $\sqsubseteq$ be the sorted order}

\For{ each $c$ in {\CutPoints} }
   \State CandInv[$c$] := $\emptyset$;
   \Comment{Set of candidate invariants at $c$}
\EndFor
\State CandInv[$\mathsf{Start}$] := $\ppre$; CandInv[$\mathsf{End}$] := $\ppost$;
   \Comment{Fixed invariants at $\mathsf{Start}$ and $\mathsf{End}$}

\infocomment{Candidate invariant generation}
\For{ each segment s from $c_1$ to $c_2$, where $c_1, c_2 \in {\CutPoints} \cup
  \{\mathsf{Start}, \mathsf{End}\}$ and $c_1 \sqsubseteq c_2$}

   \If{($s$ lies within a loop)}
      \State $\ell$[s] := loop counter of innermost nested loop containing $s$;
      \State $\mathsf{OuterLoopCtrs}$[s] := loop counters of all other outer loops containing $s$;
   \Else  \Comment{ $s$ not in any loop}
      \State $\ell$[s] := $\bot$; $\mathsf{OuterLoopCtrs}$[s] := $\emptyset$;
   \EndIf
  \State $\mathsf{ScalarVars}$[s] := $\mathcal{V} ~\cup~ \mathsf{OuterLoopCtr}$[s];
  \State CandInv[$c_2$] := CandInv[$c_2$] $\cup ~\mathsf{findHeuristicCandidateInvariants}(s, c_2, \ell[s], \mathsf{ScalarVars}[s], \mathcal{A})$;
\EndFor
   
\infocomment{Tiling and verification}
\For{each segment $s$ from $c_1$ to $c_2$}
    \If{(s lies within a loop)}
      \State CandTile[s] := $\mathsf{findHeuristicTile}(s, \ell[s], \mathsf{ScalarVars}[s], \mathcal{A})$; \label{T}
      \Comment{Candidate tile for $s$}
         \State \label{C} Check conditions T1, T2* and T3* for CandTile[s], as described in Section~\ref{sec:tiling-simple};

    \If{(not timed out) AND (T1 or T3* fail)}
    \State Re-calculate CandTile[s] using different heuristics; $\mathbf{goto}$~\ref{C};
    \EndIf
    \If{ (not timed out) AND (T2* or T3* fail) AND ($c_2 \neq \mathsf{End}$)}
     \State Re-calculate CandInv[$c_2$] using different heuristics; $\mathbf{goto}$~\ref{C};
     \EndIf
  \Else \Comment{$s$ not in any loop}
  \State
   \label{T2} Check condition T2**, as described in
    Section~\ref{sec:tiling-general};
    \If{(not timed out) AND (T2** fails) AND ($c_2 \neq \mathsf{End}$)}
     \State Re-calculate CandInv[$c_2$] using
     different heuristics; $\mathbf{goto}$~\ref{T2};
     \EndIf
  \EndIf
\EndFor
\If{ timed out }
 \Return{``Time out! Inconclusive answer!'' }
\EndIf
\State \Return{``Post-condition verified!'' }
\end{algorithmic}
\end{algorithm}

In the second phase, we generate candidate invariants at each
cut-point $c$ by considering every segment $s$ that ends at $c$.  For
each such segment $s$, we identify the loop counter $\ell[s]$
corresponding to the innermost loop in which $s$ appears, and the set
of loop counters $\mathsf{OuterLoopCtrs}[s]$ corresponding to other
loops that contain (or nest) $s$.  Note that when the program fragment
in the segment $s$ executes, the
\emph{active} loop counter that increments from one execution of $s$
to the next is $\ell[s]$.  The loop counters in
$\mathsf{OuterLoopCtrs}[s]$ can be treated similar to other scalar
variables in $\mathcal{V}$ when analyzing segment $s$.  We
would like the candidate invariants identified at different cut-points
to be of the form $\forall I \,\left(\Phi(I) \implies \Psi(\mathcal{A},
\mathcal{V}, I)\right)$, whenever possible.  We assume access to a routine
$\mathsf{findHeuristicCandidateInvariants}$ for this purpose.  Note
that the candidate invariants obtained from this routine may not
actually hold at $c_2$. In the next phase, we check using tiling
whether a candidate invariant indeed holds at a cut-point, and use
only those candidates that we are able to prove.

\begin{algorithm}[!th]
 \caption{\textsc{findHeuristicTile}($s$ : segment, $\ell$: loop counter, $\mathsf{ScalarVars}$: set of scalars, $\mathcal{A}$: set of arrays)}
 \label{alg:tiling}
\begin{algorithmic}[1]
  \scriptsize
 \State Let $c_1$ be the starting cut-point (or $\mathsf{Start}$ node) of $s$;
 \For{each array ${\A}$ updated in $s$}
   \State $\UpdIndexExprs{\A}{s}$ := $\emptyset$; 
   \For{each update of the form ${\A}[e]$ := $e'$ at location $c$ in $s$}
     \Comment{$e$ and $e'$ are arith expns}

     \State $\widehat{e}$ := $e$ in terms of $\ell$, $\mathsf{ScalarVars}$, $\mathcal{A}$ at $c_1$
     \Comment{Obtained by backward traversal from $c$ to $c_1$}

     \State $\UpdIndexExprs{\A}{s}$ := $\UpdIndexExprs{\A}{s} \cup \{\widehat{e}\}$
   \EndFor
     
   \State ${\InitTile}^{\A}(\ell, j)$ := $\mathsf{Simplify}\big(\bigvee_{e \in \UpdIndexExprs{\A}{s}} (j = e)\big)$;
   \Comment{Initial estimate of tile}
   
   \For{ each $e \in \UpdIndexExprs{\A}{s}$}
     \If{$\big({\InitTile}^{\A}(\ell, e) \wedge {\InitTile}^{\A}(\ell + k, e)
       \wedge (0 \le \ell < \ell + k < \mE_{\ell})\big)$ is satisfiable}
       \State Remove $e$ from $\UpdIndexExprs{\A}{s}$;
     \EndIf
   \EndFor

   \State ${\Tile}^{\A}(\ell, j)$ := $\mathsf{Simplify}\big(\bigvee_{e \in \UpdIndexExprs{\A}{s}} (j = e)\big)$;
   \Comment{Refined tile}
   \State \Return $\bigwedge_{\A \in \mathcal{A}}{\Tile}^{\A}(\ell, \cdot)$;
\EndFor
\end{algorithmic}
\end{algorithm}



In the third phase, we iterate over every segment $s$ between
cut-point $c_1$ and $c_2$ again, and use heuristics to identify tiles.
This is done by a routine $\mathsf{findHeuristicTile}$.  The working
of our current tiling heuristic is shown in
Algorithm~\ref{alg:tiling}.  For every array update ${\A}[e] := e'$ in
segment $s$, the heuristic traverses the control flow graph of $s$
backward until it reaches the entry point of $s$, i.e. $c_1$, to
determine the expression $e$ in terms of values of $\ell[s]$,
$\mathcal{V}$, $\mathsf{OuterLoopCtrs}[s]$ and $\mathcal{A}$ at $c_1$.
Let $\UpdIndexExprs{\A}{s}$ denote the set of such expressions for
updates to $\A$ within $s$. We identify an initial tile for $\A$ in
$s$ as ${\InitTile}^{\A}(\ell[s],
j) \equiv \bigvee_{e \in \UpdIndexExprs{\A}{s}} (j = e)$.  It may turn
out that the same array index expression appears in two or more
initial tiles after this step.  For example, in
Fig.~\ref{fig:simplecode}, we obtain ${\InitTile}^{\A}(\ell, j) \equiv
(\ell \le j \le \ell+1)$, and hence
${\InitTile}^{\A}(\ell, \ell+1) \wedge
{\InitTile}^{\A}(\ell+1, \ell+1)$ is satisfiable.  While the
conditions T1, T2 and T3 do not forbid overlapping tiles in general
(non-interference is different from non-overlapping tiles), our
current tiling heuristic avoids them by refining the initial tile
estimates.  For each expression $e$ in $\UpdIndexExprs{\A}{s}$, we
check if ${\InitTile}^{\A}(\ell[s], e) \wedge {\InitTile}^{\A}(\ell[s]
+ k, e)
\wedge (0 \le \ell[s] < \ell[s] + k < \mE_{\ell[s]})$ is satisfiable.
If so, we drop $e$ from the refined tiling predicate, denoted
$\Tile^{\A}(\ell[s], \cdot)$ in Algorithm~\ref{alg:tiling}.  This
ensures that an array index expression $e$ belongs to the tile
corresponding to the largest value of the loop counter $\ell[s]$ when
it is updated.  The procedure $\mathsf{Simplify}$ invoked in lines $7$
and $11$ of Algorithm~\ref{alg:tiling} tries to obtain a closed form
linear expression (or Boolean combination of a few linear expressions)
for $\bigvee_{e \in \UpdIndexExprs{\A}{s}} (j = e)$, if possible. In
the case of Fig.~\ref{fig:simplecode}, this gives the tile $(j
= \ell)$, which suffices for proving the quantified assertion in this
example.

Sometimes, the heuristic choice of tiling or the choice of candidate
invariants may not be good enough for the requisite checks (T1, T2*,
T2**, T3) to go through.  In such cases, Algorithm~\ref{alg:overall}
allows different heuristics to be used to update the tiles and
invariants.  In our current implementation, we do not update the
tiles, but update the set of candidate invariants by discarding
candidates that cannot be proven using our tiling-based checks.  It is
possible that the tiles and candidate invariants obtained in this
manner do not suffice to prove the assertion within a pre-defined time
limit.  In such cases, we time out and report an inconclusive answer.

\section{Implementation and Experiments}
\label{sec:experiments}
\paragraph{\bfseries Implementation:}
We have implemented the above technique in a tool called \ourtool.
The tool is built on top of the LLVM/CLANG~\cite{clang} compiler
infrastructure.  We ensure that input C programs are adapted, if
needed, to satisfy the syntactic restrictions in
Section~\ref{sec:prelims}.  The current implementation is {\em fully
  automated} for programs with non-nested loops, and can handle
programs with nested loops semi-automatically.

\noindent \emph{Generating candidate invariants:} We use a
template-based dynamic analysis tool, \daikon\cite{daikon}, for
generating \textit{candidate} invariants.  {\daikon} supports linear
invariant discovery among program variables and arrays, and reports
invariants at the entry and exit points of functions.  In order to
learn candidate quantified invariants, we transform the input program
as follows.  The sizes of all arrays in the program are changed to a
fixed small constant, and all arrays and program variables that are live are
initialized with random values.  We then insert a dummy function call
at each cut-point.  Our transformation collects all array indices that
are accessed in various segments of the program and expresses them in
terms of the corresponding loop counter(s).  Finally, it passes the
values of accessed array elements, the corresponding array index
expressions and the loop counter(s) as arguments to the dummy call, to
enable {\daikon} to infer candidate invariants among them.  The
transformed program is executed multiple times to generate traces.
\daikon~learns candidate linear invariants over the parameters passed
to the dummy calls from these traces.  Finally, we lift the candidate
invariants thus identified to quantified invariants in the natural
way.

\begin{figure}[t]
\begin{minipage}{0.45\textwidth}
\center
{\scriptsize
\begin{verbatim}
void copynswap() {
  int s, i, tmp;
  int a[s], b[s], acopy[s];
  for (i = 0; i < s; i++) { //L1
     acopy[i] = a[i];
  }
  for (i = 0; i < s; i++) { //L2
     tmp = a[i], a[i] = b[i]; b[i] = tmp;
  }
  for (i = 0; i < s; i++) {
     assert(b[i] == acopy[i]);
  }
}
\end{verbatim}
}
(a)
\end{minipage}
\begin{minipage}{0.45\textwidth}
\center
{\scriptsize
\begin{verbatim}
void dummy(int a_i, int b_i, int acopy_i, int i) { }
void copynswap() {
  int s=10, i, tmp;
  int a[s], b[s], acopy[s];
  for (i = 0; i < s; i++) {
     a[i] = rand(); b[i] = rand();
  }  
  for (i = 0; i < s; i++) { //L1
     acopy[i] = a[i];
     dummy(a[i], b[i], acopy[i], i);
  }
  for (i = 0; i < s; i++) { //L2
     tmp = a[i], a[i] = b[i]; b[i] = tmp;
  }
}
\end{verbatim}
}
(b)
\end{minipage}
\caption{(a)Input program (b) Transformed Program}
\label{fig:daikon-ex}
\end{figure}



As an example, consider the input program shown
in figure~\ref{fig:daikon-ex}(a).
The transformed program is shown in figure \ref{fig:daikon-ex}(b).
In the transformed program,
arrays $a$ and $b$ are initialized to random values.
The $dummy$ function call in loop $L1$ has four arguments
$a[i]$, $b[i]$, $acopy[i]$ and $i$.
Based on concrete traces, \daikon~initially detects the candidate
invariants $(a\_i = acopy\_i)$ and $(a\_i \neq b\_i)$ on the
parameters of the dummy function.  We lift these to obtain the candidate
quantified invariants
$\forall i. (a[i] = acopy[i])$ and $\forall i. (a[i] \neq b[i])$.  In
the subsequent analysis, we detect that $\forall i, (a[i] \neq b[i])$
cannot be proven.  This is therefore dropped from the candidate
invariants (line $24$ of Algorithm~\ref{alg:overall}), and we proceed
with $\forall i, (a[i] = acopy[i])$, which suffices to prove the
post-condition.

\noindent \emph{Tile generation and checking:}
Tiles are generated as in Algorithm~\ref{alg:tiling}.  Condition T1 is
checked using \zthree~\cite{z3}, which has good support for quantifiers.
We employ \cbmc\cite{cbmc} for implementing the checks T2*, T2** and T3*.


\begin{table*}[t!]
  \scriptsize
\begin{center}
\begin{minipage}[b]{0.45\hsize}\centering
\begin{tabular}{|c|c|c|c|c|c|}
\hline
\textsc{Benchmark} & \#L & T & S+C & B & V \\ \hline \hline
init2ipc.c & 1 & \cmark 0.5 & $\dagger$ & \cmark 0.01 & \cmark 1.0 \\ \hline
initnincr.c & 2 & \cmark 5.8 & $\dagger$ & \cmark 0.01 & \cmark 0.7 \\ \hline
evenodd.c & 1 & \cmark 0.4 & $\dagger$ & \cmark 0.01 & \cmark 0.04 \\ \hline
revrefill.c & 1 & \cmark 0.6 & $\dagger$ & \cmark 0.01 & \cmark 0.79 \\ \hline
largest.c & 1 & \cmark 0.4 & $\dagger$ & \cmark 0.01 & \cmark 0.02 \\ \hline
smallest.c & 1 & \cmark 0.4 & $\dagger$ & \cmark 0.01 & \cmark 0.02 \\ \hline
cpy.c & 1 & \cmark 0.6 & $\dagger$ & \cmark 0.01 & \cmark 2.0 \\ \hline
cpynrev.c & 2 & \cmark 3.8 & $\dagger$ & \cmark 3.1 & \cmark 5.4 \\ \hline
cpynswp.c & 2 & \cmark 4.2 & $\dagger$ & \cmark 12.4 & \cmark 1.38 \\ \hline
cpynswp2.c & 3 & \cmark 10.2 & $\dagger$ & \cmark 198 & \cmark 7.2* \\ \hline
01.c & 1 & \cmark 0.44 & $\dagger$ & \cmark 0.05 & \cmark 0.38 \\ \hline
02.c & 1 & \cmark 0.65 & $\dagger$ & \cmark 0.02 & \cmark 2.3 \\ \hline
06.c & 2 & \cmark 8.15 & $\dagger$ & \cmark 0.04 & \cmark 0.35 \\ \hline
27.c & 1 & \cmark 0.41 & $\dagger$ & \cmark 0.01 & \cmark 0.12 \\ \hline
43.c & 1 & \cmark 0.45 & $\dagger$ & \cmark 0.03 & \cmark 0.05 \\ \hline
maxinarr.c & 1 & \cmark 0.51 & $\dagger$ & \cmark 0.01 & \cmark 0.11 \\ \hline
mininarr.c & 1 & \cmark 0.53 & $\dagger$ & \cmark 0.02 & \cmark 0.13 \\ \hline
compare.c & 1 & \cmark 0.44 & $\dagger$ & \cmark 0.04 & \cmark 0.62 \\ \hline
palindrome.c & 1 & \cmark 0.52 & $\dagger$ & \cmark 0.02 & \cmark 0.39 \\ \hline
copy9.c & 9 & \cmark 34.6 & $\dagger$ & \cmark 0.46 & TO\\ \hline
init9.c & 9 & \cmark 29.2 & $\dagger$ & \cmark 0.34 & \cmark 0.16 \\ \hline
seqinit.c & 1 & \cmark 0.45 & $\dagger$ & \cmark 0.03 & \cmark 0.43\\ \hline
nec40t.c & 1 & \cmark 0.50 & $\dagger$ & \cmark 0.06 & \cmark 0.48 \\ \hline
sumarr.c & 1 & \cmark 0.55 & $\dagger$ & \cmark 0.56 & \cmark 4.2 \\ \hline
vararg.c & 1 & \cmark 0.42 & $\dagger$ & \cmark 0.03 & \cmark 0.12 \\ \hline
find.c & 1 & \cmark 0.52 & $\dagger$ & \cmark 0.02 & \cmark 0.14 \\ \hline
running.c & 1 & \cmark 0.62 & $\dagger$ & \cmark 0.04 & \cmark 0.12 \\ \hline
revcpy.c & 1 & \cmark 0.7 & $\dagger$ & \cmark 0.01 & \cmark 0.73 \\ \hline
revcpyswp.c & 2 & \cmark 6.3 & $\dagger$ & \cmark 0.02 & TO \\ \hline
revcpyswp2.c & 3 & \cmark 8.6 & $\dagger$ & \cmark 0.03 & TO \\ \hline
\end{tabular}\\
(a)
\end{minipage}
\quad
\begin{minipage}[b]{0.45\hsize}\centering
\begin{tabular}{|c|c|c|c|c|c|}
\hline
\textsc{Benchmark} & \#L & T & S+C & B & V \\ \hline \hline
copy9u.c & 9 & \xmark 0.16 & \xmark 4.48 & \xmark 0.44 & \xmark 30.8 \\ \hline
init9u.c & 9 & \xmark 0.15 & \xmark 3.77 & \xmark 0.32 & \xmark 0.14 \\ \hline
revcpyswpu.c & 2 & \xmark 0.18 & \xmark 3.11 & \xmark 0.01 & TO \\ \hline
skippedu.c & 1 & \xmark 0.81 & \xmark 2.94 & \xmark 0.02 & TO \\ \hline
mclceu.c & 1 & \textbf{?} 0.37 & \xmark 2.5 & $\star$ & $\star$ \\ \hline
poly1.c & 1 & TO & $\dagger$ & \cmark 15.7 & TO \\ \hline
poly2.c & 2 & \textbf{?} 6.44 & $\dagger$ & \textbf{?} 19.5 & TO\\ \hline
tcpy.c & 1 & \textbf{?} 0.65 & $\dagger$ & TO & \cmark 25.1 \\ \hline
skipped.c & 1 & \cmark 1.24 & $\dagger$ & TO & TO \\ \hline
rew.c & 1 & \cmark 0.48 & $\dagger$ & \cmark 0.01 & TO \\ \hline
rewrev.c & 1 & \cmark 0.39 & $\dagger$ & TO & TO \\ \hline
rewnif.c & 1 & \cmark 0.49 & $\dagger$ & \cmark 0.01 & TO \\ \hline
rewnifrev.c & 1 & \cmark 0.28 & $\dagger$ & \cmark 0.01 & TO \\ \hline
rewnifrev2.c & 1 & \cmark 0.47 & $\dagger$ & \cmark 0.01 & TO \\ \hline
pr2.c & 1 & \cmark 0.51 & $\dagger$ & TO & TO \\ \hline
pr3.c & 1 & \cmark 0.70 & $\dagger$ & TO & TO \\ \hline
pr4.c & 1 & \cmark 0.68 & $\dagger$ & TO & TO \\ \hline
pr5.c & 1 & \cmark 1.32 & $\dagger$ & TO & TO \\ \hline
pnr2.c & 1 & \cmark 0.55 & $\dagger$ & TO & TO \\ \hline
pnr3.c & 1 & \cmark 0.98 & $\dagger$ & TO & TO \\ \hline
pnr4.c & 1 & \cmark 0.86 & $\dagger$ & TO & TO \\ \hline
pnr5.c & 1 & \cmark 1.98 & $\dagger$ & TO & TO \\ \hline
mbpr2.c & 2 & \cmark 6.48 & $\dagger$ & TO & TO \\ \hline
mbpr3.c & 3 & \cmark 9.24 & $\dagger$ & TO & TO\\ \hline
mbpr4.c & 4 & \cmark 12.75 & $\dagger$ & TO & TO \\ \hline
mbpr5.c & 5 & \cmark 18.08 & $\dagger$ & TO & TO \\ \hline
nr2.c & 1-1 & \cmark 1.48* & $\dagger$ & TO & TO \\ \hline
nr3.c & 1-1 & \cmark 2.02* & $\dagger$ & TO & TO \\ \hline
nr4.c & 1-1 & \cmark 2.43* & $\dagger$ & TO & TO \\ \hline
nr5.c & 1-1 & \cmark 2.90* & $\dagger$ & TO & TO \\ \hline
\end{tabular}
\\
(b)
\end{minipage}
\end{center}
\caption{Results on selected benchmarks from
  (a) \booster~ \& \vaphor~ test-suite and (b) industrial code.
  \#L is the number loops (and sub-loops, if any) in the benchmark,
  T is \ourtool,
  S+C is \smackpluscorral, 
  B is \booster, and
  V is \vaphor.
  \cmark indicates assertion safety,
  \xmark indicates assertion violation,
  \textbf{?} indicates unknown result, and
  $\star$ indicates unsupported construct.
  All the times are in seconds.
  TO is time-out.
  * indicates
  semi-automated experiments and the corresponding execution 
  times are of the automated part. See text for explanation.
 }
\label{tab:exp-results}
\vspace{-7ex}
\end{table*}



\paragraph{\bfseries Benchmarks}
We evaluated our tool on $60$ benchmarks from the test-suites of
\booster\cite{booster} and \vaphor\cite{vaphor},
as well as on programs from an industrial code base.  The benchmarks
from \booster~ and \vaphor~ test-suites
(Table \ref{tab:exp-results}(a)) perform common array operations such
as array initialization, reverse order initialization, incrementing
array contents, finding largest and smallest elements, odd and even
elements, array comparison, array copying, swapping arrays, swapping a
reversed array, multiple swaps, and the like.  Of the $135$ benchmarks
in this test suite, $66$ benchmarks are minor variants of the
benchmarks we report. For example, there are multiple versions of
programs such as {\tt copy}, {\tt init}, {\tt copyninit}, with
different counts of sequentially composed loops.  In such cases, the
benchmark variant with the largest count is reported in the table.
Besides these, there are $22$ cases containing nested loops which can
currently be handled only semi-automatically by our implementation,
and $25$ cases with post-conditions in a form that is different from
what our tool accepts.  Hence, these results are not reported here.

Benchmarks were also taken from the industrial code of a battery
controller in a car (Table \ref{tab:exp-results}(b)).  These
benchmarks set a repetitive contiguous bunch of cells in a battery
with different values based on the guard condition that gets
satisfied.  The size of such a contiguous bunch of cells varies in
different models.  The assertion checks if the cell values are
consistent with the given specification.

All our benchmarks are within 100 lines of uncommented code.  The
programs have a variety of tiles such as $4i-4 \leq j < 4i$,
$2i-2 \leq j < 2i$, $j = size-i-1$, $j = i$ etc., with the last one
being the most common tile, where $i$ denotes the loop counter and $j$
denotes the array index accessed.
\paragraph{\bfseries Experiments}
The experiments reported here were conducted on an Intel Core i5-3320M
processor with 4 cores running at 2.6 GHz, with 4GB of memory running
Ubuntu 14.04 LTS.  A time-out of 900 seconds was set for \ourtool,
\smackpluscorral\cite{smackpluscorral}, \booster\cite{booster} and
\vaphor\cite{vaphor}.  The memory limit was set to 1GB for all the tools.
\spacer\cite{spacer} was used as the SMT solver for the Horn formulas
generated by \vaphor~ since this has been reported to perform well
with {\vaphor}.  In addition, C programs were manually converted to
mini-Java, as required by \vaphor.  Since {\smackpluscorral} is a
bounded model checker, a meaningful comparison with {\ourtool} can be
made only in cases where the benchmark violates a quantified
assertion.  In such cases, the verifier option {\tt svcomp} was used
for {\corral}.  In all other cases, we have shown a $\dagger$ in the
column for {\smackpluscorral} in Table~\ref{tab:exp-results} to
indicate that comparison is not meaningful. 

\ourtool~takes
about two seconds for verifying all single loop programs that satisfy
their assertions.  For programs containing multiple loops, $10$ random
runs of the program were used to generate candidate invariants
using \daikon.  The weak loop invariant $\Inv$, mentioned in
Section~\ref{sec:tiling}, was assumed to
be \textbf{true}.  \ourtool~took a maximum of $35$ seconds to output
the correct result for each such benchmark.  The execution time
of \ourtool~includes instrumentation for \daikon, trace generation,
execution of \daikon~on the traces for extracting candidate
invariants, translating these to {\tt assume} statements for use in
{\cbmc}, proving the reported candidate invariants and proving the
final assertion.  The execution of \daikon~and proving candidate
invariants took about $95\%$ of the total execution time.

To demonstrate the application of our technique on programs with
nested loops, we applied it to the last four benchmarks in Table
\ref{tab:exp-results}(b), each of which has a loop nested inside another.  We
used \ourtool~to automatically generate tiles for these programs.  We
manually encoded the \textit{sliced post-condition} queries and ran
{\cbmc}.  We did not have time to automate trace generation for
\daikon~and for making the above CBMC calls automatically for this
class of programs.  We are currently implementing this automation.

\paragraph{\bfseries Analysis}
\booster~and \vaphor~performed well on benchmarks from their
respective repositories.  Although \vaphor~could analyze the benchmark
for reversing an array, as well as one for copying and swapping
arrays, it could not analyze the benchmark for reverse copying and
swapping.  Since the arrays are reversed and then
swapped, all array indices need to be tracked in this case, causing
\vaphor~to fail.
\vaphor~also could not verify most of the industrial
benchmarks due to two key reasons that are not handled well by
{\vaphor}: (i) at least two distinguished array cells need to be
tracked in these benchmarks, and (ii) updates to the arrays are made
using non-sequential index values.

\booster~could analyze all the examples in which the
assertion gets violated, except for a benchmark containing an
unsupported construct (shift operator) indicated by $\star$.  This is
not surprising since finding a violating run is sometimes easier than proving an
assertion.
\booster~however could not prove several other industrial benchmarks because it
could not accelerate the expressions for indices at which the array was
being accessed.
\ourtool, on the other hand, was able to generate interesting tiles for almost all 
these benchmarks.

In our experiments,
\smackpluscorral~successfully generated counter-examples for 
all benchmarks in which the assertion was violated.  As expected, it
was unable to produce any conclusive results for benchmarks with
parametric array sizes where the quantified assertions were satisfied.

\paragraph{\bfseries Limitations}
There are several scenarios under which {\ourtool} may fail to produce
a conclusive result.  {\ourtool} uses {\cbmc} with small loop
unwinding bounds to find violating runs in programs with shallow
counter-examples.  Consequently, when there are no short
counter-examples (e.g. in {\tt mclceu.c}), {\ourtool} reports an
inconclusive answer.  {\ourtool} is also unable to report conclusively
in cases where the tile generation heuristic is unable to generate the
right tile (e.g. in {\tt tcpy.c}), when \daikon~generates weak
mid-conditions (e.g. in {\tt poly2.c}) or when {\cbmc} takes too long
to prove conditions T2* or T3* (e.g. in {\tt poly1.c}).

Our work is motivated by the need to prove quantified assertions in
programs from industrial code bases, where we observed interesting
patterns of array accesses.  Our tile generation heuristic is strongly
motivated by these patterns.  There is clearly a need to develop more
generic tile generation heuristics for larger classes of programs.

\vspace*{-0.1in}
\section{Related Work}
\vspace*{-0.1in}
\label{sec:related}
\noindent
The {\vaphor} tool~\cite{vaphor} uses an abstraction to transform
array manipulating programs to array-free Horn formulas, parameterized
by the number of array cells that are to be tracked.  The technique
relies on Horn clause solvers such
as \zthree\cite{z3}, \spacer\cite{spacer} and
\eldarica\cite{eldarica} to check the satisfiability of the generated array-free Horn
formulas.  {\vaphor} does not automatically infer the number of array
cells to be tracked to prove the assertion.  It also fails if the
updates to the array happen at non-sequential indices, as is the case
in array reverse and swap, for example.  In comparison, \ourtool~
requires no input on the number of cells to be tracked and is not
limited by sequential accesses.  The experiments in~\cite{vaphor} show
that Horn clause solvers are not always efficient on problems arising
from program verification.  To be efficient on a wide range of
verification problems, the solvers need to have a mix of heuristics.
Our work brings a novel heuristic in the mix, which may be adopted in
these solvers.

{\booster}~\cite{booster} combines acceleration
\cite{acceleration1,acceleration2} and lazy abstraction with
interpolants for arrays~\cite{lazyabsarray} for proving quantified
assertions on arrays for a class of programs.  Interpolation for
universally quantified array properties is known to be hard
\cite{Jhala,Monniaux2015}.  Hence, {\booster} fails for programs
where simple interpolants are not easily computable.  Fluid
updates~\cite{fluid} uses bracketing constraints, which are over- and
under-approximations of indices, to specify the concrete elements
being updated in an array without explicit partitioning.  This
approach is not property-directed and their generalization assumes
that a single index expression updates the array.

The analysis proposed in \cite{Gopan,Halbwachs} partitions the array into
symbolic slices and abstracts each slice with a numeric scalar
variable.  These techniques cannot easily analyze arrays with
overlapping slices, and they do not handle updates to multiple indices
in the array or to non-contiguous array partitions.  In
comparison, \ourtool~ uses state-of-the-art SMT solver
\zthree\cite{z3} with quantifier support~\cite{QuantifierZ3} for
checking interference among tiles and can handle updates to multiple
non-contiguous indices.  

Abstract interpretation based techniques \cite{Rival,ArrayCousotCL11}
propose an abstract domain which utilizes cell contents to split array
cells into groups.  In particular, the technique in~\cite{Rival} is
useful when array cells with similar properties are non-contiguously
present in the array.  All the industrial benchmarks in our test-suite
are such that this property holds.  Template-based
techniques~\cite{Gulwani} have been used to generate expressive
invariants.  However, this requires the user to supply the right
templates, which may not be easy in general.  In~\cite{transform}, a
technique to scale bounded model-checking by transforming a program
with arrays and possibly unbounded loops to an array-free and
loop-free program is presented.  This technique is not compositional,
and is precise only for a restricted class of programs.

There are some close connections between the notion of tiles as used
in this paper and similar ideas used in compilers.  For example,
tiling/patterns have been widely used in compilers for translating
loops into SIMD instructions~\cite{TreeTiler,reVectorize}.  Similarly,
the induction variable pass in LLVM can generate all accessed index
expressions for an array in terms of the loop counters.  Note,
however, that not all such expressions may be part of
a \textit{tile} (recall the tiles in Fig.~\ref{fig:simplecode}).
Hence, automatically generating the right \textit{tile} remains a
challenging problem in general.

\vspace*{-0.1in}
\section{Conclusion}
\vspace*{-0.1in}
\label{sec:conclusion}
Programs manipulating arrays are known to be hard to reason about.
The problem is further exacerbated when the programmer uses different
patterns of array accesses in different loops. In this paper, we
provided a theory of tiling that helps us decompose the reasoning
about an array into reasoning about automatically identified tiles in
the array, and then compose the results for each tile back to obtain
the overall result.  While generation of tiles is difficult in
general, we have shown that simple heuristics are often quite
effective in automatically generating tiles that work well in
practice.  Surprisingly, these simple heuristics allow us to analyze
programs that several state-of-the-art tools choke on.  Further work
is needed to identify better and varied tiles for programs
automatically.

\bibliographystyle{unsrt}
\bibliography{diff-inv}

\end{document}